\documentclass[12pt]{iopart}

\usepackage{iopams}
\usepackage{graphicx}
\usepackage{epstopdf}

\begin{document}

\title[The search for black hole binaries using a genetic algorithm]{The search for black hole binaries using a genetic algorithm}

\author{Antoine Petiteau, Yu Shang  and Stanislav Babak}

\address{Max-Planck-Institut fuer Gravitationsphysik, 
Albert-Einstein-Institut, \\
Am Muchlenberg 1, D-14476 Golm bei Potsdam, Germany}
\ead{Antoine.Petiteau@aei.mpg.de}

\begin{abstract}
In this work we use genetic algorithm to search for the gravitational wave signal from 
the inspiralling massive Black Hole binaries in the simulated LISA data.  We consider a single signal in the Gaussian instrumental noise. This is a first step in preparation for analysis of the third round
of the mock LISA data  challenge. We have extended a genetic algorithm utilizing the  properties 
of the signal and  the detector response function. The performance of this method is comparable, if not better, to already existing algorithms.

\end{abstract}

\vspace{2pc}
\maketitle

\section{Introduction}

Coalescence of two massive Black Holes (MBHs) in the galactic nuclei 
is the strongest and most promising source of gravitational waves (GW) for LISA.
This source is so powerful that the merger will be detected throughout the Universe.
The observed event rate and the recovered parameters of the binary systems will allow us to 
trace the history of MBH formation~\cite{Sesana:2007sh}. Gravitational wave 
observations should enable us to measure masses and spins of MBHs in the binary with 
unprecedented accuracy  \cite{SpinBBHLangHugues}. It is expected that
almost all the MBHs are spinning, and the value of spin might vary according to the channel 
of MBH formation. For example, if a significant part of the mass of a MBH was acquired through an accretion disk
then one can expect the spin of the MBH close to the maximum possible value.

The search for inspiralling binaries in LIGO/VIRGO data usually utilizes 
templates for non-spinning compact objects. The reason for such a choice is that non-spinning 
templates can achieve a reasonable match which are sufficient for the detection \cite{Broeck2009}. The follow up
analysis could reveal whether compact objects have spins. 
In the case of LISA, one usually does not look for a detection but rather for an accurate estimation of parameters.
As it was shown in \cite{Vecchio, SpinBBHLangHugues}, including spins should significantly improve the estimation of parameters due to the de-correlation of the parameters entering the expression of the GW phase.

Several algorithms for detecting non-spinning MBH binaries in simulated LISA data have already been demonstrated \cite{JPLCaltech, SMBHCornishPorter, MultiNest_BBH, CosmicSwarms}.
In this paper we describe a particular adaptation of a genetic algorithm to search for the GW signal from the inspirall of two MBHs.  
Genetic algorithms (GA) belong to the family of optimization methods, i.e. they look for extrema. 
The first application of GA in LISA data analysis was proposed in \cite{GenAlgoCrowder} for Galactic binaries. 
The general principle is that a waveform template is associated with an organism, and parameters play the role of the set of genes defining this organism. 
The logarithm of likelihood obtained with a given template defines the quality of the organism.
Then we evolve a set (colony) of organisms by using breeding, mutation and custom designed accelerators with the aim of finding the genotype with the highest quality. 
This corresponds to the standard Darwin's principle: ``weak dies, strong lives'', or, translated into the conventional data analysis language:
by evolving a set of templates, we are searching the parameters that maximize the likelihood.  

The work presented below is a first step toward the analysis of the third round of mock LISA data challenge. 
The mock data set consists of the Gaussian instrumental noise, Galactic background and four to six signals from the inspiralling spinning MBH binaries in a quasi-circular orbit \cite{MLDC3}.  
We made a number of simplifications: (i) we consider a single signal in the Gaussian instrumental noise (ii) we fix all the parameters of the signal except masses, sky location and time of coalescence. 
Effectively we have reduced the problem to the non-spinning case, however, the signal is quite different from the non-spinning one. 
Actually, we did not put spins equal to zero, we have fixed them, therefore the amplitude and phase modulations caused by the presence of spins are still there. 
The GA comes with a large number of options and many free parameters, and the above simplifications allow us to concentrate ourselves on  tuning the algorithm before extending it to more complicated data and signal. In addition, this enables us to estimate 
performance of the algorithm and to compare it with the alternative approaches.

The structure of the paper is as follows. 
We describe the model of the GW signal in Section~\ref{S:Model}. 
Then we give the basic principles of a standard GA in Section~\ref{S:GA}. 
In Section~\ref{S:Acc} we consider some specific modification and improvements of GA suitable for MBH binaries. 
We give and discuss our results in Section~\ref{S:Res}, and we conclude with the summary in Section~\ref{S:summary}.

\section{Template and quality estimation}
\label{S:Model}
\subsection{Model of the signal}

The waveform is described by fifteen parameters: the two masses $m_{1}$ and $m_{2}$, the initial direction of the orbital angular momentum, polar angle $\theta_{L}$ and azimuthal angle $\phi_{L}$, the initial direction of spins, polar angles $\theta_{S_{1}}$ and $\theta_{S_{2}}$ and azimuthal angles $\phi_{S_{1}}$ and $\phi_{S_{2}}$, the dimensionless spin parameters, $\chi_{1}$ and $\chi_{2}$,  the time at coalescence $t_{c}$, the phase at coalescence $\Phi_{c}$, the ecliptic coordinates of the source latitude $\beta$ and  longitude $\lambda$, and the luminosity distance $D_{L}$.

The spin-orbit and the spin-spin interactions induce a precession of each binary member's spin vector and the orbital angular momentum.  The precession equations are given in \cite{ApostolatosSpin} and the waveform used in the mock data is described in details in \cite{MLDC3}. 

We have used two orthogonal (with uncorrelated noise) TDI (time delay interferometry) channels
 A and E in the phase domain (strain). TDIs are the time delayed combinations of the measurements 
 which significantly reduces the  laser noise  \cite{VallisCsqPerteLiens, TDIRevue}. In our template, we consider a long wavelength approximation of these signals \cite{ModAmpCornish, ThesePetiteau}. This approximation ($L\omega \ll 1$, where $L$ is  armlength of LISA  and $\omega$ is an instantaneous frequency of GW)  works pretty well below approximately 5 mHz. We also have assumed rigid LISA with equal arms. Then the waveform takes the following form \cite{ModAmpCornish, ThesePetiteau}

\begin{eqnarray}
 h_{I} (t) &\simeq & 2 L \ \sin \Delta \phi_{2L}(t_{k})  \left\{ - h_{S0+}(t_{k})  \left[ \cos{(2 \psi(t_{k}) )}  F_{+I}(t_{k})   - \right. \right.\nonumber\\ 
 & &\left. \left.\sin{(2\psi(t_{k}) )}  F_{\times I}(t_{k})  \right] \sin \phi' (t_{k}) 
  + h_{S 0 \times}(t_{k})  \left[ \sin{(2 \psi(t_{k})  )}  F_{+I}(t_{k})  +
   \right. \right. \nonumber \\
 & & \left. \left. \cos {(2 \psi(t_{k})  )}  F_{\times I}(t_{k})  \right]  \cos \phi' (t_{k})  \right\},\label{TDILW}
\end{eqnarray}
where $I=\{ A, E \}$, $t_{k}$ is the time in LISA frame, $\psi$ is the polarization angle, $h_{S0+}$ and $h_{S 0 \times}$ are the polarization amplitudes  of the GW in the source reference frame, $ \Delta \phi_{2L} (t) = ( \phi (t) - \phi (t - 2L) ) / 2 $ , $ \phi' (t) = ( \phi (t) + \phi (t - 2L) ) / 2 $ with $\phi(t)$ being the phase of GW.  The antenna pattern functions $F_{+I}$ and $F_{\times I}$ have the following expression:

\begin{eqnarray}
F_{+} (\theta_{d},\lambda_{d}; t, \Omega ) & =  & {1 \over 32} \left[ 6 \sin ( 2\theta_{d} ) \left( 3 \sin\left( \Phi_{T}(t) + \lambda_{d} +  \Omega \right) -  \sin\left( 3  \Phi_{T}(t) - \lambda_{d} +  \Omega \right) \right) \right. \nonumber \\
 & &  \left. - 18 \sqrt{3} \sin^{2} \theta_{d} \sin \left( 2 \Phi_{T}(t) + \Omega \right) \right. 
 - \sqrt{3} \left( 1 + \cos^{2}\theta_{d} \right) \times
 \nonumber  \\
&  &  \left. \left( \sin \left(4  \Phi_{T}(t) - 2 \lambda_{d} + \Omega \right) + 9 \sin \left(2\lambda_{d} + \Omega \right) \right) \right] \label{TDILW_Fp} \\
F_{\times} (\theta_{d},\lambda_{d}; t, \Omega ) & =  & {1 \over 16} \left[ \sqrt{3} \cos \theta_{d} ( \cos ( 4  \Phi_{T}(t) - 2 \lambda_{d} + \Omega ) -  \right. \nonumber \\
&  & \left.   9 \cos ( 2 \lambda_{d} + \Omega ) )+  6 \sin \theta_{d} ( \cos ( 3  \Phi_{T}(t) - \lambda_{d} + \Omega ) +  \right. \\
& & \left.  3 \cos (  \Phi_{T}(t) + \lambda_{d} + \Omega ) ) \right]  \label{TDILW_Fc}
\end{eqnarray}
with $\theta_{d} = \beta + \pi/2 $, $\lambda_{d} = \lambda + \pi$, $\Phi_{T}(t) = 2 \pi t / Year$ and  $\Omega = 0$ for A and $\Omega = \pi/4$ for E.

\subsection{Quality estimation}

As we have mentioned before we associate the quality of the organism with logarithm of the likelihood. For a template $h(\widehat{\theta})$  the likelihood is given by 
\begin{equation}
\mathcal{L} (\widehat{\theta}) = C e^{- \langle s - h(\widehat{\theta}) \mid s - h(\widehat{\theta}) \rangle /2}
\label{E:Likelihood}
\end{equation}
where $s$ is the signal and $\langle h \mid s \rangle$ is the inner product defined as
\begin{equation}
\langle h \mid s \rangle  =  2 \int_{0}^{\infty} df \  {\widetilde{h} (f) \; \widetilde{s}^{*} (f) + \widetilde{h}^{*} (f) \; \widetilde{s} (f) \over S_n(f)  }
\label{E:InnerProduct}
\end{equation}
The tilde denotes the Fourier transform and $S_n(f)$ corresponds to the power spectral density of 
LISA instrumental noise (plus possible Galactic background approximated as described in \cite{CornishRubbo}).
The set of parameters $\widehat{\theta}$  maximizing the likelihood is called maximum likelihood estimator \cite{ MFFinn, Owen, CutlerFlanagan}, and that is what we are searching for.

 Before using some numerical maximization methods, it is possible to maximize over the luminosity
 distance and the phase at coalescence analytically. This is the usual procedure routinely used in the
 ground-based GW data analysis (see for example \cite{JaranowskiKrolakSchutz}). 
 
 We also efficiently maximize the likelihood over the time of coalescence by sliding the 
 template across the data (we use correlation instead of the inner product). This procedure is described  in \cite{SMBHCornishPorter, Babak}.

\section{Genetic algorithm}
\label{S:GA}

\subsection{Basic principle.}
We start with a brief introduction to the genetic algorithm, then give a detailed
implementation and describe specific modifications to solve the problems at hand. 

The GA is based on the natural selection principle which is the evolution theory established by C. Darwin. In the nature the organisms adapt themselves to their environment: the 
smartest/strongest/healthiest organism has most chances to survive and participate in 
 breeding to produce the offspring. These two processes, selection and breeding, are used in genetic algorithm for making a new generation. Since the best organisms are most likely to participate
in breeding, the new generation should be better than the previous one (at least no worse). 
 In order to increase the exploration of the possible organisms, a third process, called mutation,
  has been introduced. One needs to associate some measure of ``goodness'' with each organism. In our case it 
 is the logarithm of the likelihood $(\log \mathcal{L} )$, which we want to ``improve'' through the evolution of the 
 organisms. 

Each GW template is associated with an organism. And the set of parameters characterizing the signal, $\widehat{\theta}$, play the role of genes.  For describing genes we use a binary representation of each
parameter.   The allowed resolution for each parameter depends on the number of bits used for describing it. If the parameter $x_{k}$ has a prior range between $x_{k,min}$ and $x_{k,max}$, its resolution is $\Delta x_{k} = (x_{k,max} - x_{k,min}) / 2^{N_{k}}$, where $N_{k}$ is the number of bits. 

We keep the number of organisms in each generation fixed and allow only one active generation (one generation at the time) : it  means that after the breeding process all the parents are replaced by the children.  
Next we describe in details the three main processes : selection, breeding and mutation.  

\subsection{Selection}

The selection process chooses the parents for breeding. The probability of selecting one organism is related to its quality in the following way. First the quality $Q_{i}$, i.e. the maximized log-likelihood, of all the organisms is computed (index $i$ refers to the organism). Then we attach to each organism the
probability to be chosen for breeding as $Q_{i} / \sum_{j}^{N} Q_{j}$, and apply roulette selection method. With this method the ``good'' organisms are chosen more often than the ``bad'' ones.
Here we do not take into account the ``geographical'' proximity between parents (in other words possible correlation between templates in the same generation). By forbidding the breeding between
correlated parents we might improve efficiency to explore larger parameter space, but we reduce the overall resolution of the method, so our selection is based only on the quality.

\subsection{Breeding}
\begin{figure}[!htb] 
\center
\includegraphics[width=7cm]{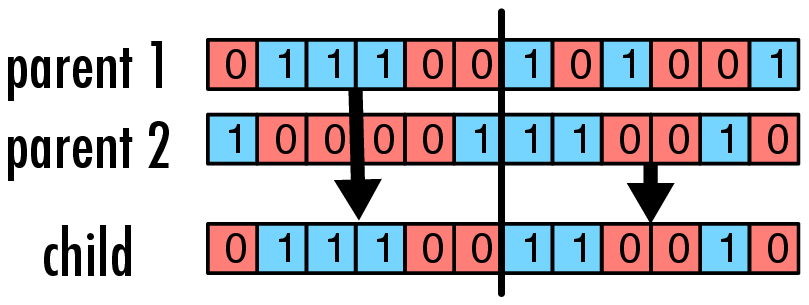}
\includegraphics[width=7cm]{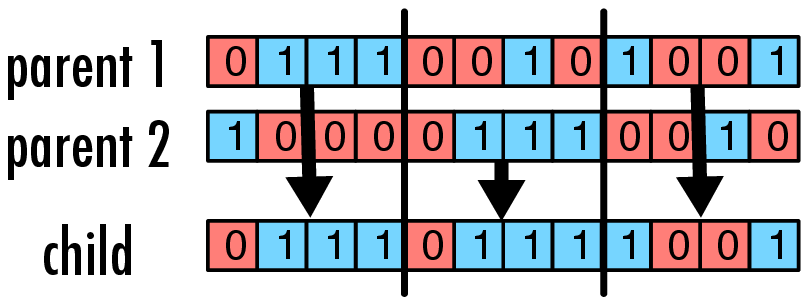}
\caption{Examples of breeding: cross-over one point on the left and cross-over two points
on the right panels} 
\label{F:BreedCO1} 
\label{F:BreedCO2} 
\end{figure}

The second process in GA is breeding. Two selected parents produce one child. The genes of the child are constructed by mixing the corresponding genes of each parent. We take one part from 
the first parent and the other part from the second parent. Depending on which parts are chosen, there are several types of breeding. We use two different types : cross-over one point (mid-point) and cross-over two points. For the cross-over one point, each gene of the child is combined out of 
the first $N/2$ bits of the first parent and last $N/2$ bits of the second parent  (see the left
panel of the Figure~\ref{F:BreedCO1}). It is similar to cross-over two points: the gene of the child is build from three equal parental parts (see the right panel of the Figure~\ref{F:BreedCO2}).

The first generation is chosen randomly by drawing parameters uniformly within the priors specified 
in \cite{MLDC3}. The chosen selection implies that the quality of our organisms more likely will increase with each generation. But, if we use only breeding, the range of resulting genes is quite restricted: it is just a combination of the parts from the initial generation. Therefore the exploration of the parameter space is rather poor. 

\subsection{Mutation}
In order to increase  the exploration of the parameter space, we need to introduce mutation. It works in 
a way similar to the mutation in the nature. Mutation is a random change of a few alleles in a gene; in our algorithm it corresponds to changing a few bits in a gene of an organism. The probability of mutation is described by a number from 0 to 1 and it is called the probability mutation rate (PMR). We mutate each gene independently and there are several types of mutation. 
First we need to decide whether we mutate a gene or not, and, if yes, we need to decide on the rule
of  mutation (how we do it).
The first possibility: we always mutate the gene and mutation is applied to each bit of gene 
independently. For each bit we draw a random number from 0 to 1 and flip the bit if the number is lower than PMR. 
The second possibility is that we decide to mutate the gene if a randomly drawn number is below
PMR. In this case we have used two different rules for mutation:
(i) we flip $N$ randomly chosen bits 
(ii) we flip $N$ adjacent bits.
Different types of mutation together with the value of PMR define the exploration area of the parameter space. Large PMR implies frequent mutations and, therefore exploration of large
parameter space, this is a desired feature in the beginning of the search. 


\section{Accelerations of genetic algorithm}
\label{S:Acc}
We have introduced above three fundamental processes of any GA. However,  
one  might need to evolve a very large number of generations before finding the best organisms.
To reduce the required number of generations and give more stability to the algorithm, we introduce several accelerators which are used in our search and described below.  

\subsection{Standard accelerators}
In this part, we describe the known accelerators often used in GA.

\subsubsection{Elitism}
The elitism (or cloning) implies that we keep the best organism through the generations. It is
 possible to clone one or several best organisms into the new generation. The elitism stabilizes the GA because the best 
 point in the parameter space  is always preserved and it guarantees the convergence of the algorithm.

\subsubsection{Simulated annealing}
The simulated annealing was already employed in LISA data analysis \cite{SMBHCornishPorter}
and proven to be very useful. The smoothness of the quality surface is controlled by the introduced temperature. If it is hot, the quality surface is very smooth and nearly all the organisms (``good'' and 
``bad'') can be selected for breeding with a similar probability. If it is cold, the quality 
surface is very peaked around the maxima and only the best organisms can be selected. 
Usually, we use a high temperature at the beginning  of the search to increase the exploration area and cool it down as we approach the solution.

\subsubsection{Evolution of PMR}
As we said above, the PMR controls the exploration of the parameter space.  Therefore we can control the search area by changing the PMR during the evolution. Usually we start with a large value for the PMR (about 0.2), then slowly decrease it and give more importance to the breeding.    
Note that we can control the exploration area using both simulated annealing and PMR. We 
use them both as they perform somewhat differently (simulated annealing uses combination of the 
initial genes without adding new) and the best result is usually achieved 
if we combine them together.

\subsection{Others accelerations}

In this part, we describe non-standard acceleration processes introduced by us and which 
utilize the properties of the signal and/or the antenna beam pattern.

\subsubsection{Brother}
We introduced what we call ``the brother'' system. With each clone we associate one organism (brother) created by using specific rules. In our application of the GA for  black hole binaries, the brother is used for searching around the mirrored (antipodal) sky position of each clone
(we keep all the parameters of the clone but flip the sky location).

\subsubsection{Local mutation}
 The binary representation creates boundaries in the parameter space. For example, the separation between the gene value 011111 and 100000 is equal to the resolution $\Delta x$  (i.e. minimal distance), but, as one can see,  it is necessary to flip all the bits for making this small step. This problem can be solved by introducing the local mutation. It corresponds to a small (of order few 
$\Delta x$) random change in the parameter value after mutation which can push it across the 
boundary.
 
\subsubsection{Fixing bits}
To find the traces of the signal in the data, we want to explore a large parameter space
at the beginning of the search. But after few hundred generations, some of the parameters are already quite
well estimated so the exploration of the large range in those parameters starts to slow down the search. At this stage we fix (freeze) the most significant bits of those parameters which reduces the allowed dynamical range.

\subsubsection{Specific breeding and mutation}
As mentioned above, different types of breeding and mutation have different properties
(main difference is in the exploration area around the best organism).
The genes (i.e. parameters) do not have the same evolution during the search. For example, the time of coalescence and the chirp mass converge to the answer quicker than other parameters. We can customize our algorithm so that each gene has its own type of breeding and mutation and which 
 can also evolve together with the colony of organisms.

\subsubsection{Change of environment}
A change of the environment corresponds to a change in the quality distribution in the parameter space. The main idea here is that the ``good'' organism remains ``good'' in the different environment.
  Here, we change the maximum allowed frequency of our template (terminate the 
inspiral at earlier times). The quality distribution for the shortened waveform has not the same 
local maxima as the quality obtained with the full waveform, but the same global maximum. Therefore, the alternating use of full and chopped waveform helps to move the search away
 from the local maxima where it has tendency to get stack.

\section{Results}
\label{S:Res}

 Our ultimate intent is to use GA in the Mock LISA Data Challenge 3~\cite{MLDC3}. 
  The mock data consist of the Gaussian stationary instrumental noise, a Galactic 
  cyclo-stationary background and four to six signals from the inspiralling spinning BHs 
in the quasi-circular orbits.

As we have mentioned in the introduction we have simplified the search with the prime purpose 
to explore the numerous options which GA provides and make the first assessment of
 the  GA performance. We consider a  single GW signal in the instrumental noise.
Further we have fixed all the parameters to their true values besides masses, sky location and time of coalescence.  The spin terms come into GW phase at 1.5 post-Newtonian (spin-orbital coupling) orders and at 2nd post-Newtonian (spin-spin) orders, so they bring rather small corrections to the phase. 
 In our simplification we want to make sure that we can find spin unrelated parameters before hunting the small corrections. We do not exclude the possibility that including all the 15 parameters in the search will bring more correlations between parameters and creating more local maxima in the likelihood which will require extension of our 
 algorithm. 
 
 Let us emphasize that the main aim of this paper is to describe the application of GA to the
 search for SMBH binaries in the LISA data, tune the algorithm by using the simplified 
 data and reduced parameter space and assess the performance of the algorithm. We do 
 not intend to conduct neither an intensive study of the performance nor the detailed comparison with other algorithms. This simplified search is a first step toward the search for precessing SMBH binaries.
 Application of GA to the full problem will be described in the separate publication \cite{FullPaper}.

 The sources can be separated into two types : the sources with a time of coalescence within the observational time and the ones which coalesce outside the observational time. We have investigated those two cases separately and we present our results below.

\subsection{Coalescence within the observational time}
\label{SS:CoalIn}
\begin{figure}[!htb] 
\center
\includegraphics[width=8cm,angle=270]{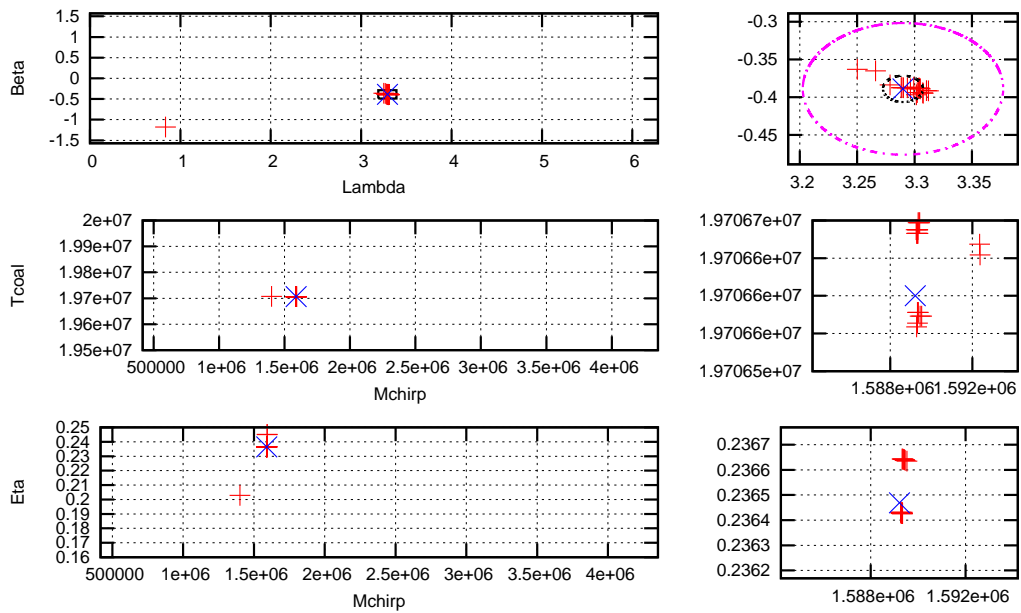}
\caption{Distribution of  50 best organisms from 50 runs of GA. The source coalesces within the observational time. The pluses '\textsf{+}' correspond to the best organism of each run after 2000 generations. The cross '\textsf{x}' corresponds to the true value of the parameters. On the left side, graphs show the full prior range of parameters. On the right side, we zoom onto the area around the 
true value. On the top-right graph, the big circle corresponds to the error of 19.63 square degrees and the small one  is 0.785 square degree error area.} 
\label{F:ResultsCoalIn} 
\end{figure}

We have analyzed  one year of simulated LISA data including the signal with following parameters:
 signal-to-noise ratio SNR = 1250, 
 $\{$ $\beta = -0.38896 \; rad$, 
$\lambda = 3.28992\; rad$, 
$ t_{c} =  19706568.3273 \; sec$, 
$M_{c} =  1589213.34 \; M_{\odot}$,  
$\eta = 0.23647$
$\}$.
We have run GA fifty times with different random initial states and for 2000 generations each 
(usually enough to get to the answer). Figure~\ref{F:ResultsCoalIn} shows the best organism obtained from each evolution. For 48 out of 50 runs, the median absolute error in the sky position is  
0.12 square degrees, the median fractional error in the chirp mass is  0.0077 \% , the median error 
in $\eta$ is 0.045\% and the median absolute error in time of coalescence is 114 sec.
Note that the errors (especially for the time of coalescence) are larger than what would be predicted 
by the Fisher matrix analysis. This is due to a large systematic error caused by using long wavelength
approximation. This systematic error is removed if we employ the rigid adiabatic approximation 
or the full response. The main result here is that we have found the global maximum.  

\subsection{Coalescence outside the observational time}
\label{SS:CoalOut}
\begin{figure}[!htb] 
\center
\includegraphics[width=8cm,angle=270]{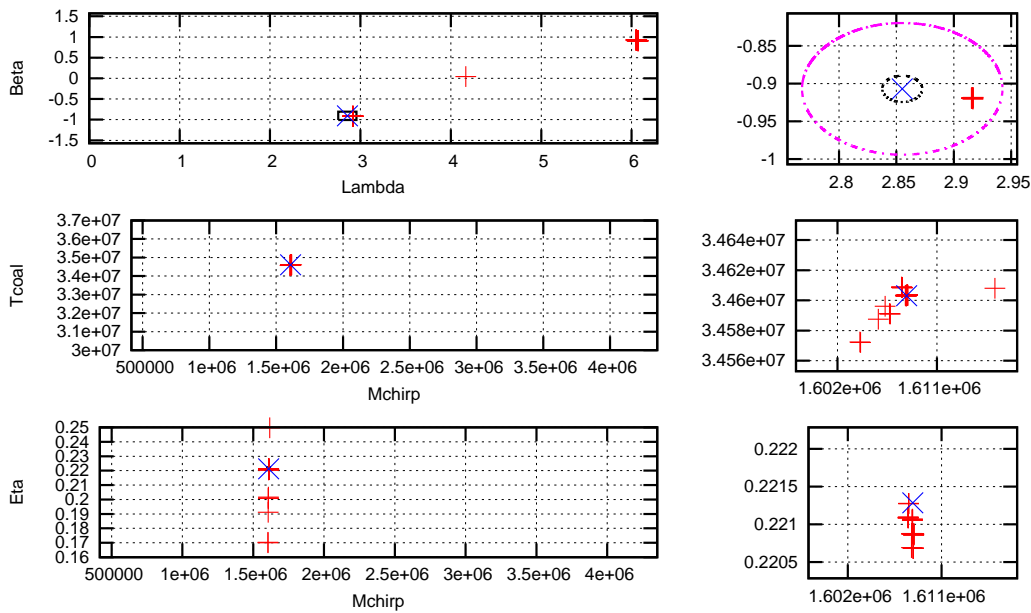}
\caption{Results of 50 runs of GA for the source with coalescence beyond the end of the observational time. The pluses '\textsf{+}' correspond to the best organisms of each run after 2000 generations. The cross '\textsf{x}' corresponds to the true value of the parameters. On the left side, graphs show the full prior range of the parameters. On the right side, we zoom around the true value. On the top-right plot, the big circle corresponds to the 19.63 square degrees error circle and the small one is for  0.785 square degree error region.} 
\label{F:ResultsCoalOut} 
\end{figure}

Similarly we have considered one year of simulated LISA data with a source coalescing outside the 
time of observation. The SNR of that signal was 47, and other parameters are
$\{$ 
$\beta = -0.90706300796 \; rad$, 
$\lambda = 2.85511464611\; rad$, 
$ t_{c} =  34603008.0 \; sec$, 
$M_{c} =  1608239.35302 \; M_{solar}$,  
$\eta = 0.22129$
$\}$. Figure~\ref{F:ResultsCoalOut} shows 50 best organisms  from each run of GA.  

The results are sligtly worse, mainly, because of the lower SNR: for 22 out the 50 jobs, the median absolute error on the sky is 2.4 square degrees, the median relative error on chirp mass is  0.0044 \% ,  $\eta$ determined with the median error   $ 0.18$ \% and the absolute median error in the time of coalescence is 153 sec. Other 28 best organisms have found the opposite sky position (there is a very strong degeneracy for this low frequency source) with similar error in the chirp mass, $\eta$ and the time at coalescence. The systematic errors here are negligible (besides systematic bias in the sky location) and those results are consistent with an analytic analysis based on the Fisher information matrix.

\section{Conclusion}
\label{S:summary}
We have presented an extended version of the GA to search for 
 BHs in the simulated LISA data. We devised several acceleration
procedures for GA which are based on the properties of the signal and the 
detector antenna pattern. The algorithm  shows remarkable performance 
for a single signal in the simulated instrumental noise assuming the precession-related 
parameters are known. The performance of GA is comparable to the best available 
algorithms based on Markov Chain Monte Carlo \cite{MLDC3, MLDC2}.
We have recently applied GA to the full 15 parameters search and obtained very good results 
which we delegate to the separate publication \cite{FullPaper}.

\section{Acknowledgement}
Work of A.P. and S.B. was supported in parts by DFG grant SFB/TR 7 
ÒGravitational Wave AstronomyÓ and by DLR (Deutsches Zentrum f¬ur Luft- und Raumfahrt). 
Y.S. was supported by MPG within the IMPRS program.

\section*{References}


\end{document}